\documentclass{iau}
\usepackage{graphicx}
\usepackage{xcolor}

\title[Impact of luminous AGN] 
{Establishing the impact of luminous AGN with multi-wavelength observations and simulations}

\author[C.M. Harrison]   
{C.M. Harrison$^1$, A. Girdhar$^{1,2,3}$ and S.R. Ward$^{2,3,4}$}

\affiliation{
$^1$School of Mathematics, Statistics and Physics, Newcastle University, U.K. \\ email: {\tt christopher.harrison@newcastle.ac.uk}\\
$^2$European Southern Observatory, Karl-Schwarzschild-Straße 2, 85748 Garching bei München, Germany\\
$^3$Ludwig Maximilian Universität, Professor-Huber-Platz 2, 80539 München, Germany\\
$^4$Excellence Cluster ORIGINS, Boltzmannstraße 2, 85748 Garching bei München, Germany
}

\pubyear{2023}
\volume{378}  
\pagerange{119--126}
\jname{Title of your IAU Symposium}
\editors{G. Bruni, M. Diaz-Trigo, K. Fukumura, S. Laha, eds.}
\begin{document}

\maketitle

\begin{abstract}
Cosmological simulations fail to reproduce realistic galaxy populations without energy injection from active galactic nuclei (AGN) into the interstellar medium (ISM) and circumgalactic medium (CGM); a process called `AGN feedback'. Consequently, observational work searches for evidence that luminous AGN impact their host galaxies. Here, we review some of this work. Multi-phase AGN outflows are common, some with potential for significant impact. Additionally, multiple feedback channels can be observed simultaneously; e.g., radio jets from `radio quiet' quasars can inject turbulence on ISM scales, and displace CGM-scale molecular gas. However, caution must be taken comparing outflows to simulations (e.g., kinetic coupling efficiencies) to infer feedback potential, due to a lack of comparable predictions. Furthermore, some work claims limited evidence for feedback because AGN live in gas-rich, star-forming galaxies. However, simulations do not predict instantaneous, global impact on molecular gas or star formation. The impact is expected to be cumulative, over multiple episodes.
\end{abstract}

\firstsection 
\section{Introduction}
There is a consensus across galaxy formation models and simulations that considerable energy injection, into the interstellar medium (ISM) and beyond, from active galactic nuclei (AGN) is required to regulate star formation and to reproduce many of the observable properties of massive galaxies (e.g., \cite[Schaye et~al. 2015]{Schaye15}; \cite[Pillepich et~al. 2018]{Pillepich18}; \cite[Dav\'{e} et~al. 2019]{Dave19}). This theoretical work has motivated a multitude of observational work, searching for `direct' evidence that AGN impact upon the properties of the ISM, or circumgalactic medium (CGM), and/or the host galaxy's star formation (see e.g., \cite[Harrison et~al. 2017]{Harrison17}). 

We provide a brief overview of some work studying the most luminous AGN (i.e., roughly with $L_{\rm AGN}>$10$^{43}$\,erg\,s$^{-1}$), which are associated with rapidly accreting supermassive black holes and that are identified mostly using optical, infrared or X-ray wavelengths. Three popular approaches that observers take to investigate if luminous AGN impact their host galaxies are discussed: (1) measuring mass outflow rates and kinetic coupling efficiencies (Section~\ref{sec:outflows}); (2) comparing the location of jets and outflows to spatially-resolved ISM properties and sites of star formation (Section~\ref{sec:spatial}); and (3) investigating the star formation rates and molecular gas properties of AGN host galaxies (Section~\ref{sec:Content}). Section~\ref{sec:conclusions} finishes this article by discussing what can be learnt about the impact of luminous AGN from these observations, within the context of galaxy formation simulations. 

\section{Outflow rates and kinetic coupling efficiencies}\label{sec:outflows}
AGN are known to drive multi-phase `outflows' into the interstellar medium (ISM) and circumgalactic medium (CGM), as evidenced by a variety of emission-line and absorption-line studies (see e.g., \cite[Cicone et al. 2018]{Cicone18}; \cite[Harrison et al. 2018]{Harrison18}; \cite[Veilleux et~al. 2020]{Veilleux20}). These AGN outflows can result in increased turbulence and/or can expel material from the host galaxy's ISM (at least temporarily).

A common approach to assess the potential impact of AGN outflows is to measure their properties, including the mass outflow rates and kinetic powers. These calculations typically involve: identifying which gas is outflowing; converting fluxes to gas masses; measuring the velocities and spatial distribution of the outflowing gas; and then applying models or basic assumptions to calculate outflow rates (\cite[Harrison et al. 2018]{Harrison18}).  

The inferred mass outflow rates in some sources are found to exceed the star formation rates (see discussion in e.g., \cite[Harrison 2017]{Harrison17}; \cite[Fiore et~al. 2017]{Fiore17}). This means that gas is being removed more rapidly than it is forming stars, implying the potential for a significant impact on the host galaxy. However, it is very challenging to infer the long-term impact of such outflows, including understanding what fraction of this material ultimately escapes the host galaxy. Furthermore, these extreme rates are not ubiquitous. The inferred outflow rates of some more `typical' systems can be very low to negligible (e.g., \cite[Davies et~al. 2020]{Davies20}; \cite[Ramos Almeida et~al. 2022]{ramosAlmeida22}; \cite[Villar Martin et~al. 2023]{VillarMartin23}). There are also analysis challenges, with very different results possible when applying different assumptions or using different gas tracers (e.g., \cite[Harrison et~al. 2018]{Harrison18}; \cite[Davies et~al. 2020]{Davies20}). 

Another common approach in the literature is to measure the ratio of outflow's kinetic power to AGN luminosity to derive the  `kinetic coupling efficiency' (see e.g., \cite[Fiore et~al. 2017]{Fiore17}). In some cases, these measurements have been compared to simulations, which invoke AGN feedback to establish if the observations are consistent -- or not -- with feedback models. However, these comparisons are very challenging. It is important to note that the observations are typically measuring an outflow rate in one particular gas phase and often without information on how the properties vary spatially. Furthermore, the AGN luminosities are `instantaneous' and do not contain information about earlier times, including when the outflows may have been initially launched. Therefore, these measured kinetic coupling efficiencies are not comparable quantities to the AGN coupling efficiencies invoked in sub-grid models of simulations (see discussion in \cite[Harrison et~al. 2018]{Harrison18}). Indeed, if simulations track the kinetic powers of outflows as they propagate through the ISM, they can be significantly lower than their initial energy, and it remains extremely challenging to simulate the phases of outflows that are captured by observations (e.g., \cite[Costa et~al. 2020]{Costa20}). 

\vspace{-0.3cm}
\section{Spatially-resolved investigation of impact}

\label{sec:spatial}
\begin{figure}
    \centering
    \includegraphics[width=0.9\textwidth]{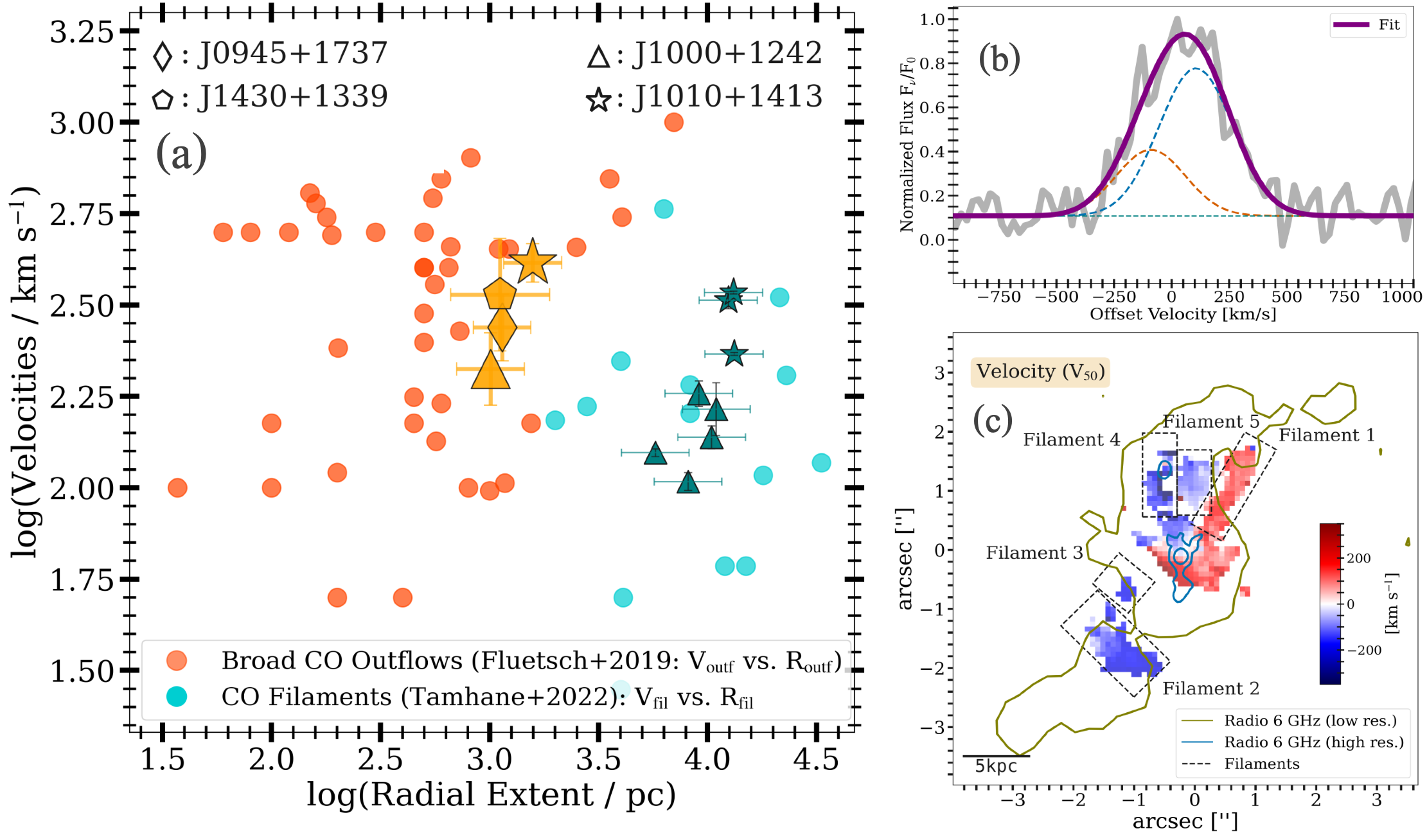}
    \caption{{\em (a)} Molecular gas velocities and radial extents for two types of AGN-ISM interaction, traced via CO emission: (1) broad emission-line components for a sample of AGN and star-forming galaxies (\cite[Fluetsch et~al. 2019]{Fluetsch2019}): (2) filaments around radio lobes for BCGs (\cite[Tamhane et~al. 2022]{Tamhane2022}). The larger symbols represent data for four $z<0.2$ `radio quiet' quasars, two of which exhibit both types of structure. {\em (b)} Example CO emission-line profile for one of these quasars. {\em (c)} Example CO velocity map of the same quasar, with radio contours, and highlighting CO structures associated with $\approx$10\,kpc radio lobes. Figures adapted from Girdhar et~al. (in prep).}
    \label{fig:girdhar}
\end{figure}

Recent numerical simulations have highlighted that multiple feedback `modes' can exist simultaneously, caused by AGN-driven winds, jets and/or radiation pressure. This can include: injected turbulence to reduce star formation efficiency; removal of high-density nuclear gas (i.e., star-forming material); halting accretion from the halo (i.e., regulating future star formation); and even the localised boosting of star formation due to shocks compressing the gas to higher densities (e.g., \cite[Costa et~al. 2020]{Costa20}; \cite[Mandal et~al. 2021]{Mandal21}). 

Observations now also show that AGN can impact upon the ISM, and wider-scale environment, in multiple different ways. For example, luminous `radio quiet' quasars, which have considerable radiative output and are often associated with powerful winds, are also able to interact with the star-forming material in their host galaxies via low-power radio jets and lobes (e.g., \cite[Girdhar et~al. 2022]{Girdhar22}; \cite[Audibert et~al. 2023]{Audibert23}; \cite[Cresci et~al. 2023]{Cresci23}). An example of this is showcased in Figure~\ref{fig:girdhar}, where broad, and asymmetric CO emission-line profiles (tracing molecular outflows in the inner regions of galaxies) and molecular filaments entrained by expanding radio lobes are observed simultaneously (also see \cite[Morganti et~al. 2023]{Morganti23}). These processes are typically searched for asynchronously in observations of luminous AGN (e.g., \cite[Fluetsch et~al. 2019]{Fluetsch}) and brightest cluster galaxies (BCGs; e.g., \cite[Russel et~al. 2019]{Russel19}; \cite[Tamhane et~al. 2022]{Tamhane22}), respectively. 

The aforementioned observations are, at least qualitatively, similar to expectations of simulations of individual galaxies, which predict AGN will ultimately have a negative impact upon a host galaxy (e.g., \cite[Costa et~al. 2020]{Costa20}; \cite[Mandal et~al. 2021]{Mandal21}). However, this global, and long-term suppression of star formation, may not easily be associated with a single AGN episode, or a localised ISM interaction. This could help explain why `direct' observational evidence of star formation suppression continues to be scarce, tentative or inconclusive (e.g., \cite[Scholtz et~al. 2021]{Scholtz21}; \cite[Cresci et~al. 2023]{Cresci23}). On the other hand, observations that show depleted and/or excited molecular gas within the ISM at the locations of low-power jets or outflows can indicate that the star-forming material is directly affected by the AGN (e.g., \cite[Rosario et~al. 2019]{Rosario19}; \cite[Girdhar et~al. 2022]{Girdhar22}; \cite[Audibert et~al. 2023]{Audibert23}). Nonetheless, the {\em long-term} and {\em global, galaxy-wide impact} of such processes remains more speculative. This motivates a statistical approach, as described below, that measures the global molecular gas and star formation properties of large samples of AGN host galaxies.

\section{Gas content and star formation rates}
\label{sec:Content}

\begin{figure}
    \centering
    \includegraphics[width=0.5\textwidth]{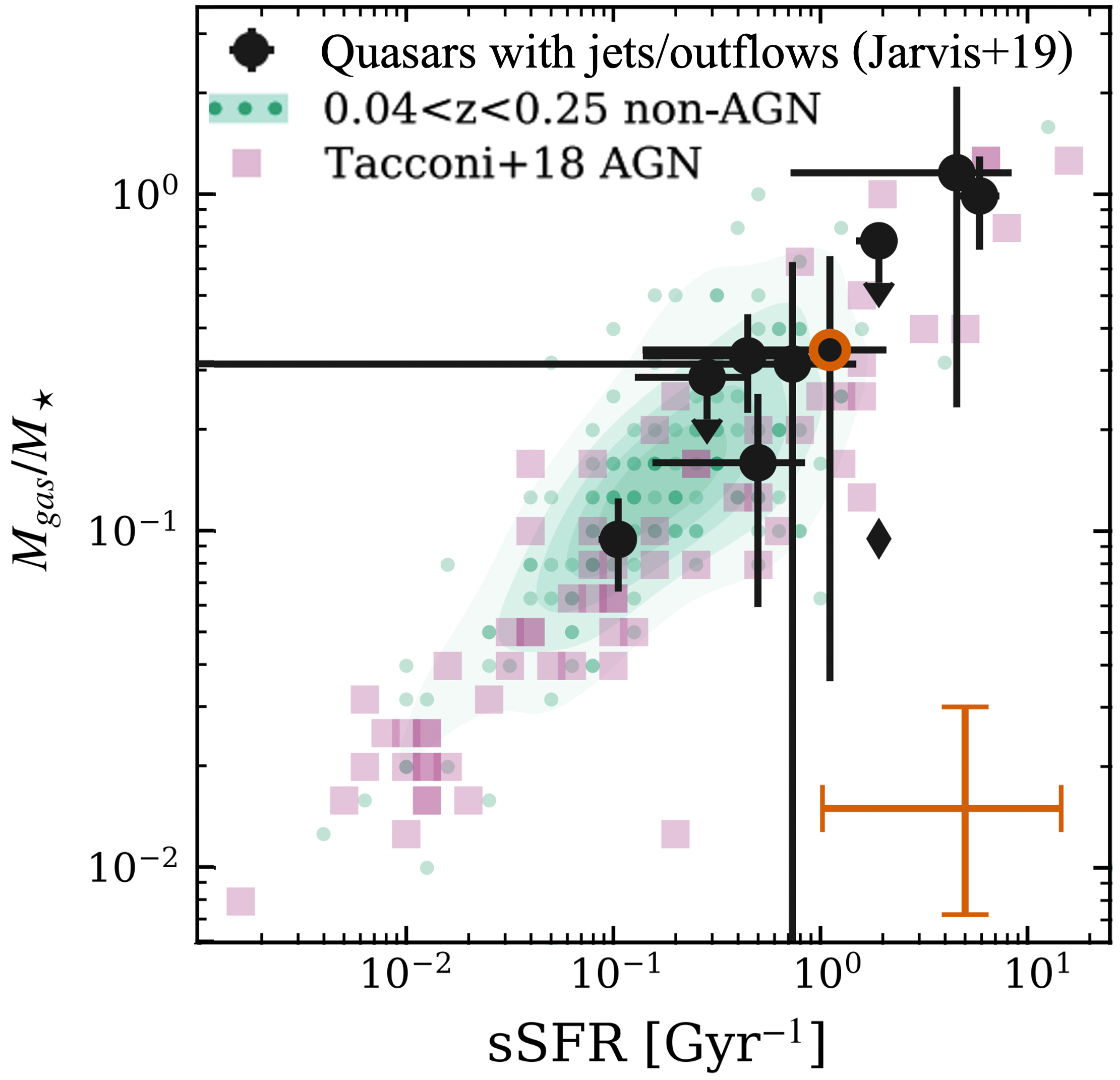}
    \caption{Ratio of molecular gas to stellar mass versus specific star formation rate. Galaxies from \cite[Tacconi et~al. (2018)]{Tacconi18} with and without an identified AGN are represented by density contours (also green points) and by magenta squares, respectively. Type 2 $z<0.2$ quasars, with known outflows and/or jets are shown with black circles. Adapted from \cite[Jarvis et~al. (2020)]{Jarvis20}.}
    \label{fig:jarvis20}
\end{figure}

One popular approach to investigate the impact of AGN on their host galaxies, is to take galaxy samples and measure how the star formation properties and/or the molecular gas properties (i.e., the `fuel' for star formation) are related (or not) to the presence of an AGN. These studies come in various forms, including: (1) investigating trends of star formation rates or molecular gas content with AGN luminosities (e.g., \cite[Stanley et~al. 2015]{Stanley15}; \cite[Ramasawmy et~al. 2019]{Ramasawmy19}; \cite[Shangguan et~al. 2020]{Shangguan20}); and (2) establishing if AGN host galaxies have different distributions of star formation and/or molecular gas properties to non-AGN host galaxies (e.g., \cite[Scholtz et~al. 2018]{Scholtz18}; \cite[Koss et~al. 2021]{Koss21}; \cite[Valentino et~al. 2021]{Valentino21}). 

One specific example of this type of analysis is shown in Figure~\ref{fig:jarvis20}. The molecular gas fractions (derived from CO observations) are plotted as a function of specific star formation rate (i.e., star formation rate divided by stellar mass) for low redshift galaxies and AGN. The star-forming galaxies and AGN from \cite[Tacconi et~al. (2018)]{Tacconi18} are shown to occupy the same area of parameter space. Furthermore, quasars with known radio jets and powerful ionised outflows also have the same molecular gas and star formation properties than their less active galaxy counterparts (\cite[Jarvis et~al. 2020]{Jarvis20}).

With only a few exceptions (e.g., \cite[Circosta et~al. 2021]{Circosta21}; \cite[Bischetti et~al. 2021]{Bischetti21}, which investigate molecular gas content at $z\sim2$), studies of AGN across multiple epochs and using a variety of samples and approaches tend towards a consensus that luminous AGN typically live in gas-rich, star-forming galaxies. It can be tempting to conclude that this is in conflict with predictions that luminous AGN regulate star formation (as suggested by some observational studies), however, as we demonstrate below, this empirical result is not conflicting with predictions of cosmological simulations including AGN feedback.

\vspace{-0.3cm}
\section{Discussion and conclusions}
\label{sec:conclusions}

\begin{figure}
    \centering
    \includegraphics[width=0.7\textwidth]{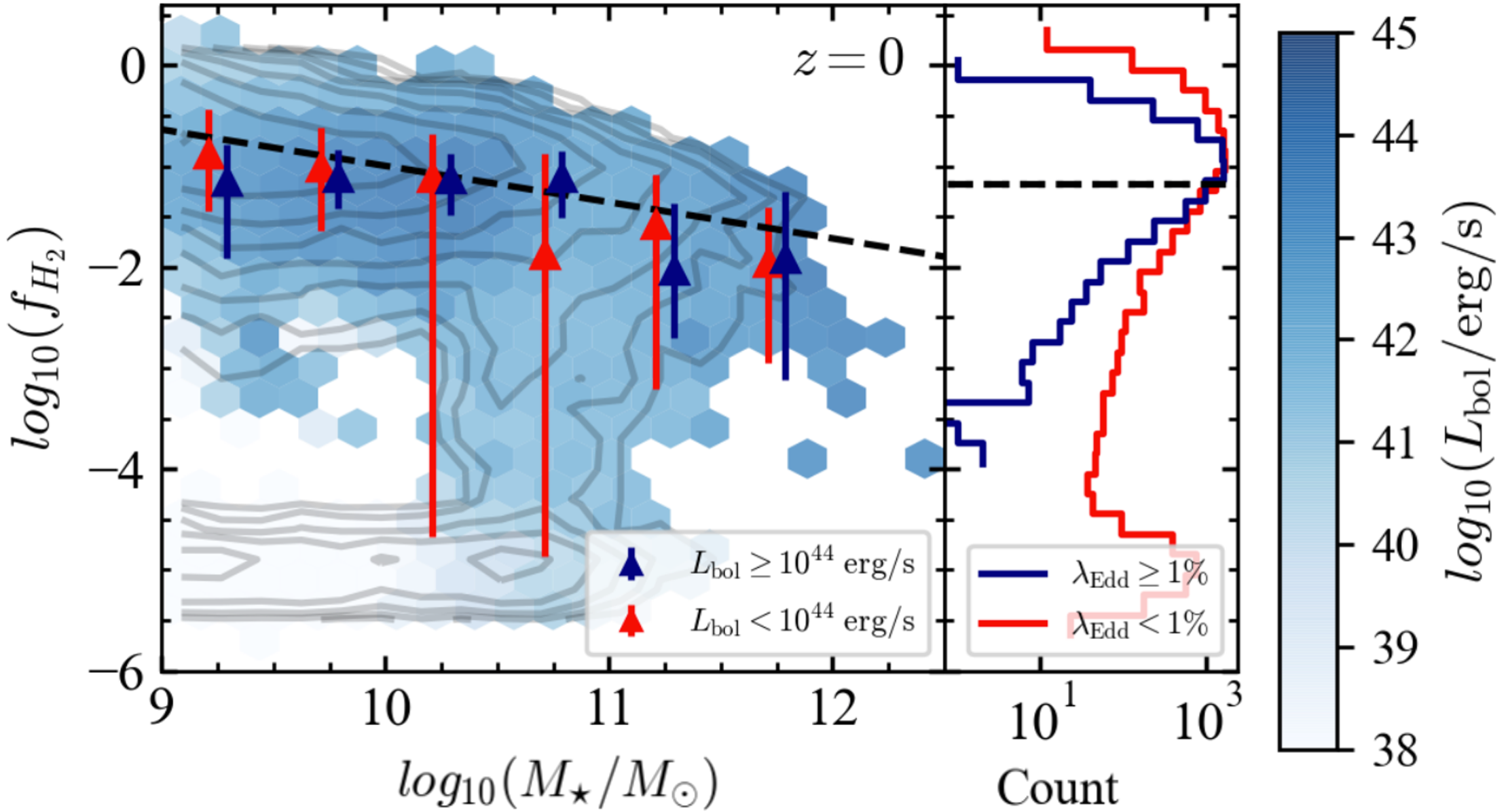}
    \caption{Ratio of molecular gas to stellar mass versus stellar mass, for $z=0$ galaxies in the IllustrisTNG100 simulation. Colouring represents the mean AGN bolometric luminosity within each bin. Contour lines represent number density, and the triangles show the median values in bins of stellar mass for galaxies classified as AGN (blue) and non-AGN (red), based on their luminosity.  The dashed line is the observed main sequence from \cite[Tacconi et~al. (2018)]{Tacconi18}. The histograms show the distributions of gas fractions for sources classified as AGN (blue) or non-AGN (red), based on their Eddington ratios. Adapted from \cite[Ward et~al. (2022)]{ward22}.}
    \label{fig:ward22}
\end{figure}

We have given a very brief overview of some of the observational work investigating the impact of luminous AGN on their host galaxies (in particular the molecular gas content and the star formation). On the one hand, there is overwhelming evidence that AGN are able to modify the distribution and properties of the ISM and CGM of their host galaxies. This is due to the ever-growing body of observations that AGN radiation, winds and/or jets are able to influence the molecular gas by injecting turbulence, driving it away, and/or exciting it (Section~\ref{sec:outflows} and \ref{sec:spatial}). On the other hand, the {\em long-term} and {\em global significance} of these processes is less well-established, with the possibility that their impact is very localised, lasts only on short timescales, and/or is insignificant. Indeed, the observation that luminous AGN tend to live in gas-rich, star-forming galaxies has sometimes been presented as evidence against effective feedback (Section~\ref{sec:Content}).

With this in mind, it is worth revisiting the cosmological simulations, which invoke AGN feedback, and motivated much of the observational work. Whilst these simulations typically lack the resolution to include a detailed physical prescription of AGN feedback, they are useful for comparing to the statistical studies of AGN and the molecular gas and star formation properties presented in Section~\ref{sec:Content}. For example, \cite[Ward et~al. (2022)]{Ward22} investigated predictions from the cosmological simulations: (1) IllustrisTNG100 (\cite[Pillepich et~al. 2018]{Pillepich18}); (2) EAGLE (\cite[Schaye et~al. 2015]{Schaye15}) and (3) SIMBA (\cite[Dav{\'e} et~al. 2019]{Dave19}). These are broadly similar in scope, with $\sim$100\,Mpc$^{3}$ boxes and simulation outputs that enable predictions of AGN luminosities, stellar masses, star formation rates and molecular gas content as a function of cosmic time. \cite[Ward et~al. (2022)]{Ward22} used these to reproduce the type of experiments performed by observers, as discussed in Section~\ref{sec:Content}. One example of this analysis is presented in Figure~\ref{fig:ward22}, where the molecular gas fractions of the $z=0$ galaxies from IllustrisTNG are presented as a function of stellar mass. The luminous AGN (selected on either luminosity or Eddington-ratio) are found to be preferentially located in gas rich galaxies. Broadly speaking, across all three simulations investigated, and at both $z=2$ and $z=0$, AGN are predicted to live in gas-rich, star-forming galaxies, in qualitative agreement with most observational work, although it is worth noting that there are distinct {\em quantitative} differences in the predictions across the three simulations due to the different feedback models used (\cite[Ward et~al. 2022]{Ward22}).

We conclude that we may not expect to have a `smoking gun' observational signature that luminous AGN have a long-term, significant impact on molecular gas content or star formation. The reasons for this are many, with some discussion on this presented in e.g., \cite[Scholtz et~al. (2018)]{Scholtz18}; \cite[Ward et~al. (2022)]{Ward22}; \cite[Piotrowska et al. (2022)]{Piotrowska22}. For example, an AGN may stop being visible {\em before} they have had an appreciable impact on their host galaxies. Furthermore, the significant impact, which ultimately results in a `quenched' galaxy, may not be attributed to a single luminous AGN episode, but instead due to the cumulative effect of many AGN episodes. Indeed, according to \cite[Piotrowska et al. (2022)]{Piotrowska22}, the strongest predictor of a galaxy as having quenched star formation (or not), is considered to be the black hole mass, rather than AGN luminosity, in both cosmological simulations and observations. In this context, black hole mass can be considered as an indirect tracer of the integrated power output of the AGN across the history of the galaxy.

In summary, the current body of evidence does not appear to be in conflict with AGN feedback models. However, work is still needed to test the different assumed models of AGN feedback. This can be done by comparing the quantitative predictions of the distributions of properties of galaxy populations from the different cosmological simulations (see e.g., \cite[Ward et~al. 2022]{Ward22}). Furthermore, the current and forthcoming generation of high-resolution individual galaxy simulations should make it possible to extract meaningful predictions of spatially-resolved outflow properties in different phases (e.g., mass outflow rates, kinetic powers), compare these to observations, and to assess the ultimate impact that they can have on galaxy evolution.

\vspace{0.3cm} \noindent CMH acknowledges a United Kingdom Research and Innovation grant (MR/V022830/1). SRW acknowledges the Deutsche Forschungsgemeinschaft (DFG, German Research Foundation) under Germany's Excellence Strategy (EXC-2094-390783311).

\end{document}